 \documentclass[prb,twocolumn,showpacs,showkeys]{revtex4}

\usepackage{graphicx}% Include figure files
\usepackage{amssymb,amsmath}
\usepackage{dcolumn}% Align table columns on decimal point
\usepackage{bm}% bold math
\usepackage{color}

\begin{document}

\title{On the metallic conductivity of the delafossites 
       $ {\bf PdCoO_2} $ and $ {\bf PtCoO_2} $}

\author{Volker Eyert}
\email{eyert@physik.uni-augsburg.de}
\altaffiliation[Permanent address: ]
               {Center for Electronic Correlations and Magnetism,
                Institut f\"ur Physik, Universit\"at Augsburg, 
                86135 Augsburg, Germany}
\affiliation{Laboratoire CRISMAT, UMR CNRS-ENSICAEN(ISMRA) 6508, 6 Boulevard
             Mar\'echal Juin, 14050 Caen Cedex, France}
\author{Raymond Fr\'esard}
\affiliation{Laboratoire CRISMAT, UMR CNRS-ENSICAEN(ISMRA) 6508, 6 Boulevard
             Mar\'echal Juin, 14050 Caen Cedex, France}
\author{Antoine Maignan}
\affiliation{Laboratoire CRISMAT, UMR CNRS-ENSICAEN(ISMRA) 6508, 6 Boulevard
             Mar\'echal Juin, 14050 Caen Cedex, France}

\date{\today}

\begin{abstract}
The origin of the quasi two-dimensional behavior of $ {\rm PdCoO_2} $ 
and $ {\rm PtCoO_2} $ is investigated by means of electronic structure 
calculations. They are performed using density functional theory in 
the generalized gradient approximation as well as the new full-potential 
augmented spherical wave method. We show that the electric conductivity 
is carried almost exclusively by the in-plane Pd (Pt) $ d $ orbitals. 
In contrast, the insulating $ {\rm CoO_2} $ sandwich layers of octahedrally 
coordinated Co atoms may be regarded as charge carrier reservoirs. 
This leads to a weak electronic coupling of the Pd (Pt) layers. The 
obtained nearly cylindrical Fermi surface causes the strong anisotropy 
of the electric conductivity. 
\end{abstract}



\maketitle

\section{Introduction}
\label{intro}

Transition metal oxides attract a lot of attention due to a great 
variety of physical phenomena, most of which go along with the 
ordering of some microscopic degrees of freedom as a function 
of, e.g., temperature, pressure, or doping. Prominent examples 
are the striking metal-insulator transitions of the vanadates  
\cite{imada98}, high-$ {\rm T_c} $ superconductivity in the cuprates, 
or the colossal magnetoresistance observed in the manganates 
\cite{helmholt93,tomioka95,raveau95,maignan95}. Cobaltates have 
aroused much interest due to the occurrence of different spin states 
\cite{eyert04,fresard04,wu05}. In addition, they are promising materials 
for thermoelectric applications \cite{terasaki97,masset00}. 

Known since 1873, when Friedel discovered the mineral $ {\rm CuFeO_2} $, 
the delafossites $ {\rm ABO_2} $ keep on generating a strong and ever 
increasing interest \cite{dupont71,tanaka98,marquardt06}, especially 
after Kawazoe {\em et al.}\ evidenced simultaneous transparency and 
p-type conductivity \cite{kawazoe97}, which laid ground for the 
development of transparent optoelectronic devices. Furthermore, the 
quasi two-dimensionality of the lattice and the triangular coordination 
of atoms gave rise to such exciting physical properties as strong 
anisotropies of the electrical conductivity and magnetic frustration 
effects. 

\begin{figure}[htb]
\centering
\includegraphics[width=0.72\columnwidth,clip]{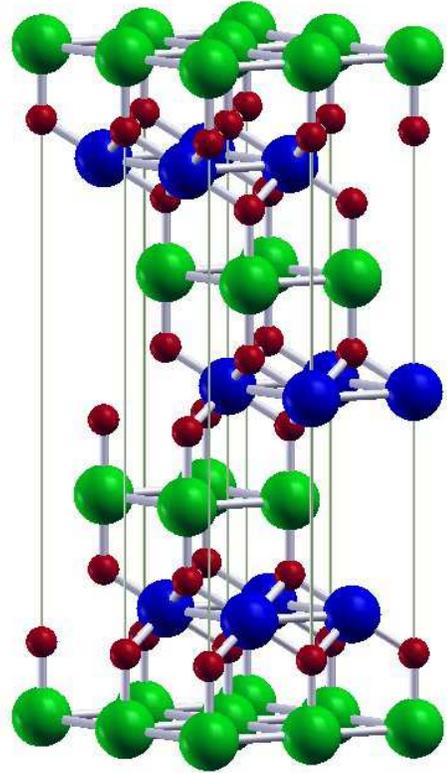}
\caption{(Color online) Crystal structure of $ {\rm PdCoO_2} $. Palladium, 
         cobalt, and oxygen atoms are shown as green, blue, and red 
         (light, big dark, and small dark) spheres, respectively.}
\label{fig1}
\end{figure}

The delafossite structure has the space group $ {\rm R\bar{3}m} $ and 
results from a stacking of monoatomic triangular layers, see Fig.\ 
\ref{fig1} \cite{dupont71,marquardt06}.
In particular, the B-atoms are at the centers of distorted oxygen 
octahedra, which share edges and form the characteristic $ {\rm BO_2} $ 
sandwich layers. These trilayers are interlinked by linear O--A--O 
bonds, resulting in a twofold coordination of the A-atoms. However, 
the latter have, in addition, six in-plane nearest neighbour A-atoms.  
For this reason, the structure may be likewise regarded as formed 
from single A-atom layers, which are intertwined by the octahedral 
sandwiches. For $ {\rm PdCoO_2} $ and $ {\rm PtCoO_2} $, we will find 
this latter point of view particularly useful. Finally, the oxygen 
atoms are tetrahedrally coordinated by one A-atom and three B-atoms. 
Pressure studies reveal an increase of the structural anisotropy on 
compression indicating the high mechanical stability of both the 
octahedral sandwich layers and the O--Pd--O (O--Pt--O) dumbbells 
\cite{hasegawa03}.  

In general, interest in the delafossite-type compounds has concentrated 
quite much on the triangular arrangement of the transition-metal atoms 
and the resulting possible frustration effects, which arise once 
localized magnetic moments are established. While most of these oxides 
have been found to be antiferromagnetic semiconductors, other class 
members like $ {\rm PdCrO_2} $, $ {\rm PdCoO_2} $, $ {\rm PdRhO_2} $, and 
$ {\rm PtCoO_2} $ attracted interest due to their rather high metallic 
conductivity. 
In particular, $ {\rm PdCoO_2} $ has been shown to possess one of the lowest 
electric resistivities of normal-state oxides, even lower than that of 
Pd metal at room temperature \cite{dupont71,tanaka98,hasegawa02}. Yet, 
the conductivity is strongly anisotropic \cite{dupont71,hasegawa02}.  
In particular, the ratio of the resistivities parallel and perpendicular 
to the $ c $ axis can be as large as $ 200 $ in $ {\rm PdCoO_2} $ 
\cite{hasegawa02}. Photoemission data indicate that the density of states 
at the Fermi energy can be exclusively attributed to the Pd $ 4d $ states 
\cite{tanaka98,higuchi98,marquardt06}. From the combination of photoemission 
spectroscopy and inverse photoemission spectroscopy several authors 
concluded that the Fermi energy is located at a shallow minimum of the 
density of states and doping might thus lead to rather high values of the 
thermoelectric power \cite{higuchi98,hasegawa02,higuchi04}.

Despite their simple chemical formulae the delafossites may be regarded 
as prototypical superlattices where the composition of both the A and 
B layers can be used to strongly influence the behavior of the whole 
system. For instance, in $ {\rm CuCrO_2} $, the Fermi energy falls into 
the Cr $ 3d $ band, but since the Cr layers order magnetically this 
compound is a magnetic semiconductor. In contrast, as will be shown 
below, in $ {\rm PdCoO_2} $, the Co layers only act a charge reservoirs, 
and conduction takes place almost exclusively in the Pd layers. 

As a matter of fact, quite a few electronic structure calculations 
for delafossite compounds have been reported in the literature 
\cite{mattheiss93,galakhov97,seshadri98,ingram01,nie02,okabe03a,kandpal02,ong07,mazin07,singh07}. 
For $ {\rm PdCoO_2} $, there exist linear muffin-tin orbitals calculations 
by Seshadri {\em et al.}\ as well as by Okabe {\em et al.}\ 
\cite{seshadri98,okabe03a}. While according to the former authors, who 
also investigated $ {\rm PtCoO_2} $, the density of states at 
$ {\rm E_F} $ is mainly due to the Pd $ 4d $ states with only small 
contributions from the Co $ 3d $ and O $ 2p $ orbitals, the results 
by Okabe {\em et al.}\ are not very specific in this respect. For this 
reason, the role of these orbitals for the metallic conductivity is 
not yet completely clear. In order to resolve the issue and to make 
closer connection with the photoemission data, we apply in the present 
work the new full-potential augmented spherical wave method to study 
the electronic properties of the title compounds. We concentrate 
especially on the strong anisotropies and on the influence of the 
different species and orbitals on the electronic properties.

\section{Theoretical Method}
\label{method}

The calculations are based on density-functional theory and the
generalized gradient approximation (GGA) \cite{pbe96} with the 
local-density approximation parametrized according to Perdew and 
Wang \cite{perdew92}. They were performed using the scalar-relativistic 
implementation of the augmented spherical wave (ASW) method (see Refs.\ 
\onlinecite{wkg,aswrev,aswbook} and references therein).
In the ASW method, the wave function is expanded in atom-centered
augmented spherical waves, which are Hankel functions and numerical
solutions of Schr\"odinger's equation, respectively, outside and inside
the so-called augmentation spheres. In order to optimize the basis set,
additional augmented spherical waves were placed at carefully selected
interstitial sites. The choice of these sites as well as the augmentation
radii were automatically determined using the sphere-geometry optimization
algorithm \cite{sgo}. Self-consistency was achieved by a highly efficient
algorithm for convergence acceleration \cite{mixpap}. The Brillouin zone
integrations were performed using the linear tetrahedron method with up
to 1469 {\bf k}-points within the irreducible wedge \cite{bloechl94,aswbook}.  

In the present work, we used a new full-potential version of the ASW
method, which was implemented only very recently \cite{fpasw}. 
In this version, the electron density and related quantities are given
by a spherical-harmonics expansion inside the muffin-tin spheres.
In the remaining interstitial region, a representation in terms of
atom-centered Hankel functions is used \cite{msm88}. However, in
contrast to previous related implementations, we here get away without
needing a so-called multiple-$ \kappa $ basis set, which fact allows
for a very high computational speed of the resulting scheme.

\section{Results and Discussion}
\label{results}

While the previous calculations were based on the crystal structure 
data by Prewitt {\em et al.}\ \cite{dupont71}, we here calculated 
these parameters from a minimization of the total energy. To this 
end, in a first step for each compound the lattice was relaxed and 
after that the oxygen parameter was optimized. The results for both 
compounds are summarized in Tab.\ \ref{tab1}. 
\begin{table}
\caption{Experimental and calculated lattice parameters (in \AA) and 
         atomic positions.}
\begin{ruledtabular}
\begin{tabular}{llccc}
compound          &        & a      &    c    & $ z_O $ \\
\hline
$ {\rm PdCoO_2} $ & exp.\  & 2.8300 & 17.743  & 0.1112 \\
                  & calc.\ & 2.8767 & 17.7019 & 0.1100 \\
$ {\rm PtCoO_2} $ & exp.\  & 2.8300 & 17.837  & 0.1140 \\
                  & calc.\ & 2.8989 & 17.458  & 0.1128 \\
\end{tabular}
\end{ruledtabular}
\label{tab1}
\end{table}
Note that the deviation of the calculated structural parameters from 
the measured ones is 2.4\% at most, which is very well within the 
known limits of the GGA and an excellent proof of the validity of the 
new full-potential ASW method. 

\begin{figure}[htb]
\centering
\includegraphics[width=0.8\columnwidth]{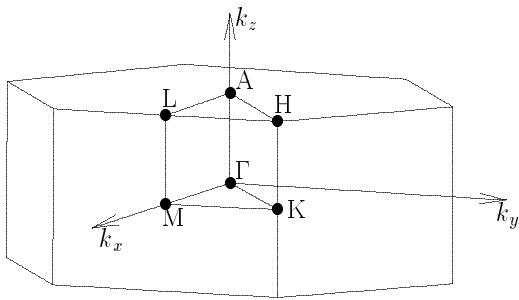}
\caption{First Brillouin zone of the hexagonal lattice.}
\label{fig2}
\end{figure}

\begin{figure}[htb]
\centering
\includegraphics[width=\columnwidth,clip]{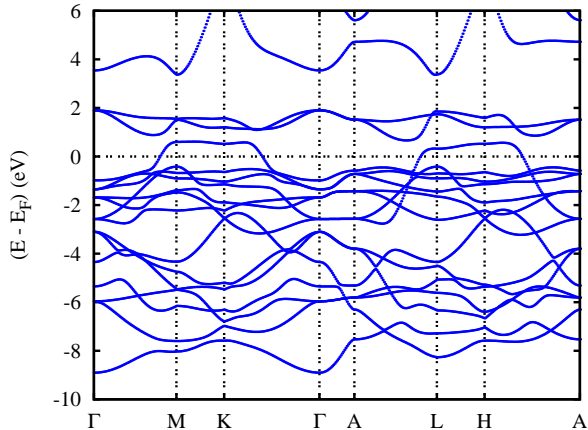}
\caption{(Color online) Electronic bands of $ {\rm PdCoO_2} $.} 
\label{fig3}
\end{figure}

\begin{figure}[htb]
\centering
\includegraphics[width=\columnwidth,clip]{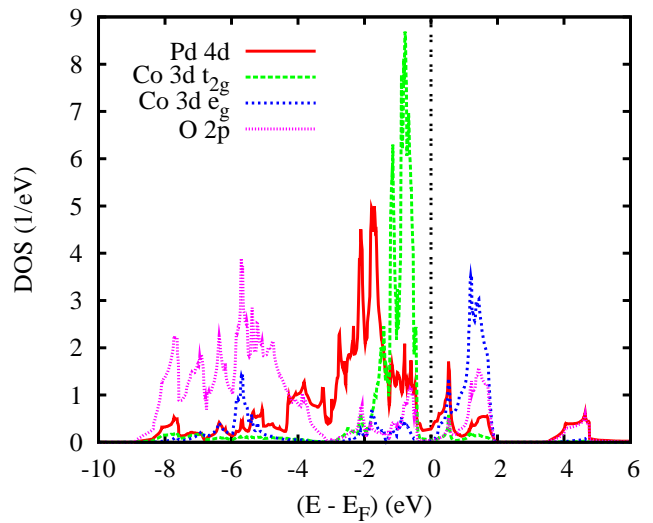}
\caption{(Color online) Partial densities of states (DOS) of 
         $ {\rm PdCoO_2} $.
         Selection of Co $ 3d $ orbitals is relative to the local 
         rotated reference frame, see text.}
\label{fig4}
\end{figure}

\begin{figure}[htb]
\centering
\includegraphics[width=\columnwidth,clip]{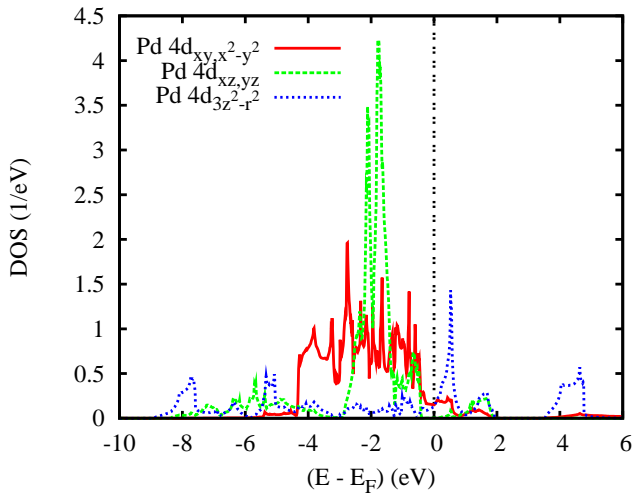}
\caption{(Color online) Partial Pd $ 4d $ DOS of $ {\rm PdCoO_2} $.} 
\label{fig5}
\end{figure}

In order to discuss the electronic properties, we first concentrate 
on $ {\rm PdCoO_2} $ and only after that discuss the changes coming 
with the substitution of Pt for Pd. 
The electronic bands along selected high-symmetry lines of the first 
Brillouin zone of the hexagonal lattice, Fig.\ \ref{fig2}, 
are displayed in Fig.\ \ref{fig3}.
The corresponding partial densities of states (DOS) are shown in 
Fig.\ \ref{fig4}.
The rather complicated structure of both the electronic bands and the 
DOS results from the energetical overlap of the relevant orbitals in 
the energy interval shown. While the lower part of the spectrum is 
dominated by O $ 2p $ states, the transition metal $ d $ states lead 
to rather sharp peaks in the interval from $ -3.5 $\,eV to $ +2 $\,eV. 
In particular, we recognize the $ t_{2g} $ and $ e_g $ manifolds of the 
Co $ 3d $ states as resulting from the octahedral coordination. In 
representing these partial DOS we have used a local rotated coordinate 
system with the Cartesian axes pointing towards the oxygen atoms. 
$ \sigma $-type overlap of the O $ 2p $ states with the Co $ 3d $ $ e_g $ 
orbitals leads to the rather sharp contribution of the latter near 
$ -5.8 $\,eV. In contrast, due to the much weaker $ \pi $-type overlap 
of the O $ 2p $ states with the $ t_{2g} $ orbitals, these states give 
rise to sharp peaks in the interval from $ -2 $\,eV to $ {\rm E_F} $. 
Since the Fermi energy falls right between the $ t_{2g} $ and $ e_g $ 
manifolds, Co turns out to be in a $ d^6 $ low-spin state. Our results 
thus provide further support to the picture that Co is trivalent while Pd is 
in a monovalent $ d^9 $ configuration \cite{dupont71,tanaka98,marquardt06}.  
Furthermore, the Co and O states give only a tiny contribution to the 
electrical conductivity, which is maintained almost exclusively by the 
Pd $ 4d $ states. The latter are further analyzed in Fig.\ \ref{fig5},
which displays the five Pd $ 4d $ partial DOS. Since Pd is linearly 
coordinated by two oxygen atoms parallel to the $ c $ axis and has 
six Pd neighbours in the $ a $-$ b $ plane we used the global coordinate 
system to represent these partial DOS. With this choice, contributions 
from the $ d_{xy} $ and $ d_{x^2-y^2} $ as well as from the $ d_{xz} $ 
and $ d_{yz} $ states are identical. The Pd $ 4d $ partial DOS are 
strongly influenced by the aforementioned coordination. In particular, 
the six peaks of the $ d_{3z^2-r^2} $ states near $ -8.0 $, $ - 6.2 $, 
$ -5.2 $, $ +0.6 $, $ 1.5 $, and $ +4.2 $\,eV are complemented by 
contributions of the O $ 2p $ partial DOS of similar shape reflecting 
the strong $ \sigma $-type $ d $-$ p $ overlap along the $ c $ axis. 
In contrast, the short in-plane Pd-Pd distances of about 2.83\AA\ 
(experimental value), which are only by 3\% longer than those in 
metallic Pd, lead to the broad Pd $ d_{xy,x^2-y^2} $ bands visible 
in the interval from $ -4.5 $ to $ 2 $\,eV in Fig.~\ref{fig5}. In 
addition, these bands give rise to the largest contribution at 
$ {\rm E_F} $, whereas the $ d_{xz,yz} $ states do not contribute 
at all. In contrast, the contribution from the $ d_{3z^2-r^2} $ 
orbitals is rather similar to that of the $ d_{xy,x^2-y^2} $ orbitals. 
However, according to a $ {\bf k} $-resolved analysis, their weights 
vary on the Fermi surface. At about $ + 0.6 $\,eV, the contribution 
of the $ d_{xy,x^2-y^2} $ orbitals is strongly suppressed, and the 
small dispersion along the line M-K causes the sharp $ d_{3z^2-r^2} $ 
peak seen in Fig.\ \ref{fig5}.

\begin{figure}[htb]
\centering
\includegraphics[width=\columnwidth,clip]{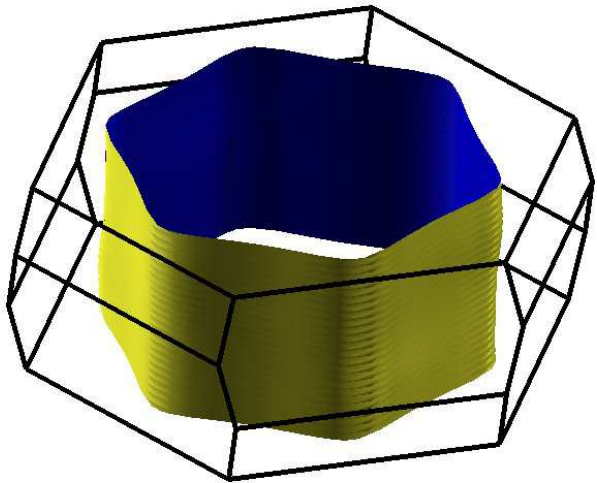}
\caption{(Color online) Fermi surface of $ {\rm PdCoO_2} $.}
\label{fig8}
\end{figure}

The strong quasi two-dimensionality of the electronic states is reflected 
by the Fermi surface depicted in Fig.\ \ref{fig8}.
Apart from the very small bending parallel to the $ c $ direction, it 
gives rise to completely in-plane Fermi velocities and explains the 
strong anisotropy in electric conductivity. However, note that, 
according to the electronic bands shown in Fig.\ \ref{fig3}, the 
dispersion parallel to the direction $ \Gamma $-A in general is not 
negligible. 
In passing, we mention the negligible dispersion along M-K, which is 
typical of tight-binding bands in a triangular lattice. These bands 
give rise to the sharp peak at about $ +0.6 $\,eV in Fig.\ \ref{fig5} 
and can thus be attributed to the Pd $ d_{3z^2-r^2} $ states. 

Overall, our results are in good agreement with the previous 
calculations by Seshadri {\em et al.}\ and by Okabe {\em et al.}\ 
\cite{seshadri98,okabe03a}. In particular, Seshadri {\em et al.}\ 
obtained a distribution of states at $ {\rm E_F} $, which is similar 
to ours. Furthermore, our results are in agreement with the 
photoemission data by Tanaka {\em et al.}\ and Higuchi {\em et al.}\ 
\cite{tanaka98,higuchi98}, who attribute the metallic conductivity 
almost exclusively to the Pd $ 4d $ states. It has been 
argued by several authors, that the dominant contribution is due 
to the Pd $ d_{3z^2-r^2} $ orbitals, which hybridize with the Pd $ 5s $ 
states. While we find indeed a $ 5s $ contribution of the order of 
$ 0.02 $\,states/eV, hence, of the order of 10\% of the $ d_{3z^2-r^2} $ 
partial DOS at $ {\rm E_F} $, we still point to the in-plane 
$ d_{xy} $ and $ d_{x^2-y^2} $ states, which play an even greater role 
at $ {\rm E_F} $ than the $ d_{3z^2-r^2} $ states. As a consequence, 
our results demonstrate that the metallic conductivity is maintained 
by the in-plane $ d_{xy} $ and $ d_{x^2-y^2} $ orbitals and the in-plane 
part of the $ d_{3z^2-r^2} $ orbitals to a similar degree with a somewhat 
greater influence of the former. This finding is confirmed by the 
calculations for $ {\rm PtCoO_2} $, where the DOS at $ {\rm E_F} $ is 
again almost exclusively due to the in-plane states. 

\begin{figure}[htb]
\centering
\includegraphics[width=\columnwidth,clip]{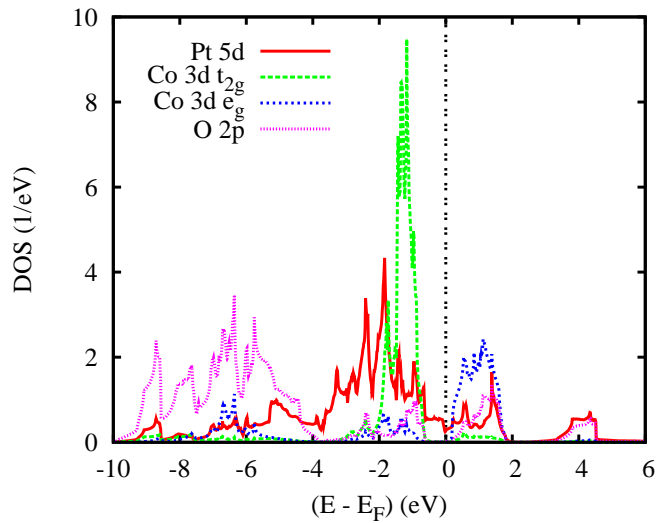}
\caption{(Color online) Partial densities of states (DOS) of 
         $ {\rm PtCoO_2} $.
         Selection of Co $ 3d $ orbitals is relative to the local 
         rotated reference frame, see text.}
\label{fig9}
\end{figure}

In general, the partial DOS obtained for $ {\rm PtCoO_2} $, which are 
displayed in Fig.\ \ref{fig9},
resemble those of the palladium system. However, the larger extent 
of the Pt $ 5d $ orbitals as compared to the Pd $ 4d $ states leads 
to an increased width mainly of the in-plane $ d $ bands as well as 
an increased overlap of these states with the O $ 2p $ states. As a 
consequence, the anisotropy of the electrical conductivity will be 
somewhat reduced in $ {\rm PtCoO_2} $ as is indeed observed  
\cite{dupont71}. 

In passing, we mention additional GGA+U calculations for 
$ {\rm PdCoO_2} $, which however, confirmed the GGA results without 
any noticeable changes.

\section{Summary}
\label{summary}

In summary, we have shown that the strongly anisotropic metallic 
conductivity of $ {\rm PdCoO_2} $ and $ {\rm PtCoO_2} $ is almost 
exclusively due to the Pd (Pt) layers. In contrast, the octahedrally 
coordinated $ {\rm CoO_2} $ sandwiches are insulating. As Co was found in 
a low-spin $ d^6 $ configuration, these $ {\rm CoO_2} $ complexes strongly 
suppress the electronic coupling between the Pd (Pt) metallic layers, and 
the $ {\rm Co^{3+}} $ layers merely act as charge reservoirs. In addition, 
they pronounce 
the quasi two-dimensionality of the system. It may be speculated how 
the introduction of impurities into the cobalt layers, i.e.\ replacing, 
e.g.\ Fe or Ni for Co, alters the electronic properties locally and 
might thus be used to design nano-structured materials. Work along 
this line is in progress. In this context one might ask to what extent 
the structural changes induced by the introduction of impurities with 
different covalent radii as compared to Co might be damped by the 
intermediate oxygen layers. This would provide a situation, where 
chemical substitution only acts on the electronic degrees of freedom.

\section{Acknowledgements}
We gratefully acknowledge many fruitful discussions with T.\ Kopp, 
C.\ Martin, and W.\ C.\ Sheets. This work was supported by the 
Deutsche Forschungsgemeinschaft through SFB 484.
Figs.\ \ref{fig1} and \ref{fig8} were generated using the XCrysDen 
software (Ref.\ \onlinecite{kokalj03}).

%
%  References
%

\end{document}